\documentclass{webofc}
\usepackage{graphicx} 
\graphicspath{{figures/}}
\usepackage{subcaption}
\usepackage[varg]{txfonts}
\usepackage[utf8]{inputenc}

\usepackage{listings}
\usepackage{xcolor}

\usepackage{natbib}
\usepackage{hyperref}
\usepackage{enumitem}

\usepackage{lineno}

\newcommand{\kt}{${k}_\text{T}$}
\newcommand{\akt}{anti-${k}_\text{T}$}
\newcommand{\Akt}{Anti-${k}_\text{T}$}
\newcommand{\JR}{\texttt{JetReconstruction.jl}}
\newcommand{\ee}{$e^+e^-$}

\title{Fast Jet Finding in Julia}

\author{\firstname{Graeme Andrew} \lastname{Stewart}\inst{1}\fnsep\thanks{\email{graeme.andrew.stewart@cern.ch}} \and
\firstname{Sanmay} \lastname{Ganguly}\inst{2} \and
\firstname{Sattwamo} \lastname{Ghosh}\inst{3} \and
\firstname{Philippe} \lastname{Gras}\inst{4}\fnsep \and
\firstname{Atell} \lastname{Krasnopolski}\inst{5}
}

\institute{CERN, Esplanade des Particules 1, Geneva, Switzerland
\and
Indian Institute of Technology, Kanpur, India
\and 
Indian Institute of Science Education and Research, Kolkata, India
\and
IRFU, CEA, Université Paris-Saclay, Gif-sur-Yvette, France
\and
Julius-Maximilians-Universität Würzburg, Würzburg, Germany
}

\abstract{%
Jet reconstruction remains a critical task in the analysis of data from HEP
colliders. We describe in this paper a new, highly performant, Julia package for
jet reconstruction, \JR, which integrates into the growing ecosystem of Julia
packages for HEP. With this package users can run sequential reconstruction
algorithms for jets. In particular, for LHC events, the \Akt, Cambridge/Aachen
and Inclusive-\kt\ algorithms can be used. For FCCee studies the use of
alternative algorithms such as the Generalised \kt\ for \ee\ and Durham are also
supported.

The performance of the core algorithms is better than Fastjet's C++
implementation, for typical LHC and FCCee events, thanks to the Julia compiler's
exploitation of single-instruction-multiple-data (SIMD), as well as ergonomic
compact data layouts.

The full reconstruction history is made available, allowing inclusive and
exclusive jets to be retrieved. The package also provides the means to visualise
the reconstruction. Substructure algorithms have been added that allow advanced
analysis techniques to be employed. The package can read event data from EDM4hep
files and reconstruct jets from these directly, opening the door to FCCee and
other future collider studies in Julia.}

\begin{document}

\maketitle

\section{Introduction}
\label{sec:introduction}

High energy physics (HEP) software is inherently
multi-lingual~\cite{pivarski2022}. Across the field actively used codes exist in
many different languages. For the code that is used in mainline HEP workflows
two languages have dominated in the last few decades: C++ and Python. C++ saw
adoption at the BaBar experiment at SLAC and, subsequently, at the Large Hadron
Collider (LHC) experiments. Python has grown enormously in popularity in recent
years, becoming almost ubiquitous across the field. These languages have
different strengths, with C++ excelling at runtime performance, used heavily for
simulation and reconstruction; and Python shining in the areas of rapid
turnaround, prototyping and steering, being then particularly strong in the
analysis domain.

The current status of C++ and Python, with these languages having rather
different paradigms, leads to friction and potentially awkward interfaces. Code
developed in Python may not run efficiently at scale, leading to inefficient use
of computing resources, even necessitating a rewrite in C++. The current
generation of physicists is generally far more comfortable in Python and there
is a loss of skills in C++, which is a challenge for the experiments.

An alternative option, which is attracting increasing interest, is to use a
language that can bring the runtime advantages of C++, but the ergonomic
advantages of Python. The \emph{Julia Programming
Language}~\cite{bib:julia_freshapproach,10.1145/3276490} was designed
specifically to do this efficiently and effectively, and has been adopted by
many scientific communities~\cite{perkel-julia-science}. In HEP, explorations of
Julia have been promising~\cite{Stanitzki:2020bnx,eschle2023potential}. In
particular, a recent comparison of Julia, Python and C++ for the task of
\emph{sequential jet finding}~\cite{polyglot-jets-chep23} found that Julia
performed as well as, or better, than C++, with improved code ergonomics. 

In this paper we report on the developments that have happened in the Julia code
presented in~\cite{polyglot-jets-chep23}, in particular the improvements that
have resulted in the recent release of the production Julia package,
\JR~\cite{jetreconstruction-jl-github}. The package has been made more
accessible to users, with comprehensive documentation and consistent interfaces.
New algorithms have been introduced targeting jet reconstruction
at \ee\ experiments, including reading data from Key4HEP's
EDM4hep~\cite{Gaede:2022leb} data format files. Support for jet substructure
analysis at $pp$ colliders has been introduced. We also give the latest
benchmarking results, that continue to demonstrate better performance than
Fastjet~\cite{Cacciari:2011ma} for almost all parameters.

\section{Production Release of Julia Jet Reconstruction}
\label{sec:prodrel}

Details of the algorithms and strategies used for $pp$ events in the Julia
version of jet finding have been described in~\cite{polyglot-jets-chep23}.
However, due to the different development history of how the two strategies,
\texttt{N2Plain} and \texttt{N2Tiled}, were implemented, the original return
values from the reconstruction were different. In the \texttt{N2Plain} case an
implicit $p_T$ cut, selecting inclusive jets, was made; whereas in the
\texttt{N2Tiled} strategy a dedicated  object that encapsulates the
reconstruction sequence was returned that contains a record of all jet merging
steps, with parents and metric distance ($d_{ij}$) parameter, allowing the
complete graph of the process to be obtained. This structure, a
\texttt{ClusterSequence} object, is much more useful as this information is
required for further jet processing. This object is now used in all sequential
reconstruction algorithms (but always through accessors, to hide internal
implementation details).

This choice then allowed the implementation of the other core jet selection,
viz.\ \emph{inclusive jets}. An interface was added where the same method can be
used to make a selection on either the number of final jets, or on the maximum
value of the metric distance ($d_{ij}$) -- this takes advantage of the fact that
in Julia method parameters can be named, providing a clearer interface than the
type based method selection in C++ (e.g.,
\texttt{inclusive\_jets(cs::ClusterSequence; ptmin=5.0)} for a $p_T$ cut).

In addition to the jet selection, another algorithm for $pp$ jet reconstruction
was added, the generalised \kt\ algorithm. This algorithm uses as a momentum
metric $k^{2p}_\text{T}$, where the power value $p$ is arbitrary (and for
specific integer powers of $-1$, $0$ and $1$, maps to the well known \akt,
Cambridge/Aachen, and inclusive-\kt\ algorithms,
respectively~\cite{Cacciari:2005hq,Matteo_Cacciari_2008,fastjetmanual}).

\begin{figure}[h]
    \begin{center}
        \includegraphics[width=0.6\linewidth]{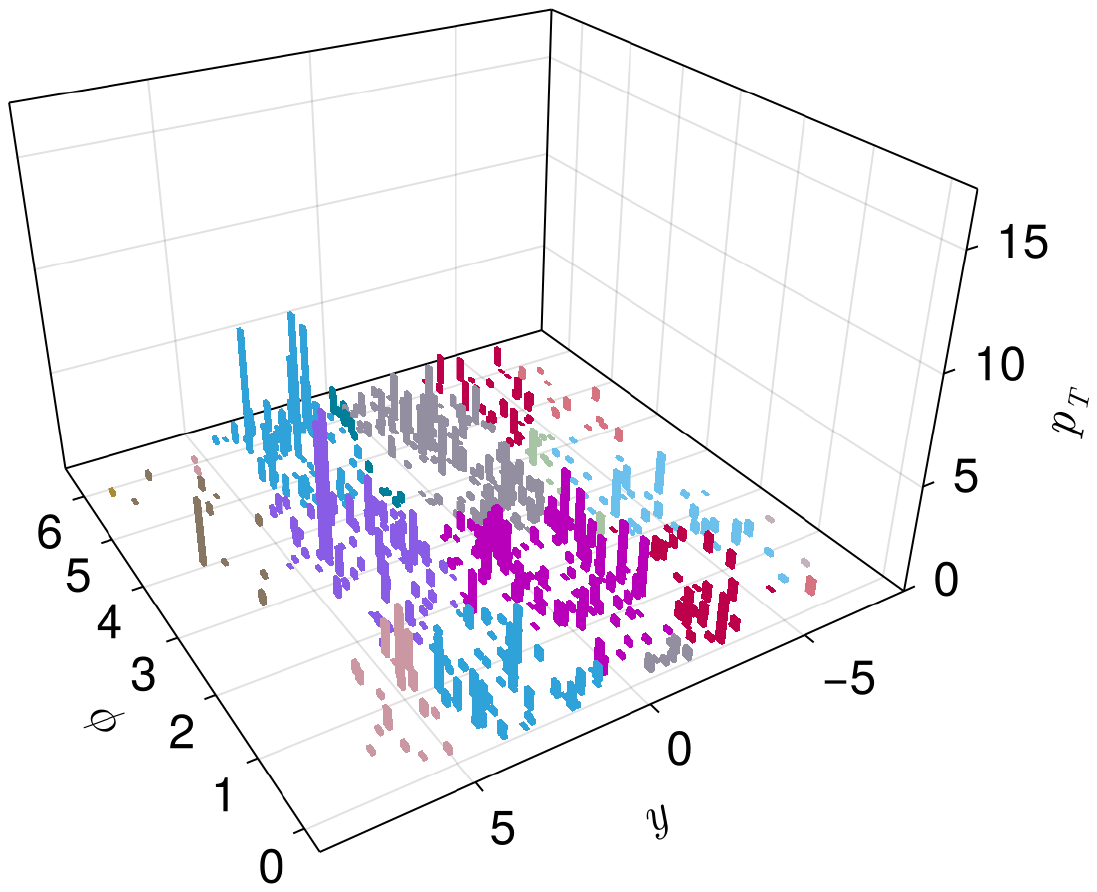}
        \caption{Visualisation of a typical $pp$ collision jet reconstruction, using \akt\ with $R=2$, in the $y-\phi$ (rapidity, azimuthal angle) plane. The height of each bar indicates the original cluster energy and the colour represents the final jet clustering, i.e., all clusters with the same colour are clustered together.}
        \label{fig:jetvisplot}
    \end{center}
\end{figure}

Support for the visualisation of jet reconstruction, through the Julia
visualisation package \emph{Makie}~\cite{Danisch2021}, has been improved. As
Makie is a heavy dependency, we take advantage of the \emph{extensions} feature
of the Julia packaging system, where the visualisation extensions to \JR\ are
only loaded if \texttt{Makie} already exists in the current user environment. An
example of the output from the visualisation extension is shown in Figure
\ref{fig:jetvisplot}. Taking advantage of the fact that all reconstruction steps
are captured by the \texttt{ClusterSequence} a new visualisation option was
added, which animates the reconstruction
process~\cite{jetrecoAnimationCHEP2024}.

Of the two core strategies, \texttt{N2Tiled} scales better to higher
initial cluster densities, as found in typical LHC $pp$ events. However, there
is an overhead for this tiling, which makes the \texttt{N2Plain} strategy better
at low cluster densities. It is highly desirable that the user would not have to
manually select a strategy, so a heuristic performance scan was made, indicating
that the performance of each strategy is about the same for 80 input clusters.
So a third strategy, \texttt{Best} was introduced, which selects
\texttt{N2Plain} for $<=80$ clusters, otherwise \texttt{N2Tiled}.

Before a useful release of the software could be made, documentation for the
package had to be written. This was done using the standard Julia documentation
support package, \texttt{Documenter.jl}~\cite{documenter-jl}, which has the
great advantage of using the inline code docstrings to document methods.
Documentation was then published onto the JuliaHEP organisation's GitHub Pages
website~\cite{jetreco-docs}.

With all of this refactoring done, and with an enhanced suite of tests added,
the first public release of \JR\, v0.3.0, was made in June 2024. The package was
added to the Julia public registry, making installing it for any user as simple
as \texttt{add JetReconstruction} from the standard Julia package interface.

\section{Support for \ee\ algorithms}
\label{sec:ee}

To add support for the reconstruction of jets in \ee\ events, some different
algorithms are needed. The essential idea of sequential jet reconstruction
remains the same: calculation of a distance metric between all clusters, then
merging the clusters with the lowest metric and repeating. However, for \ee\
events it is preferable to reconstruct in geometric space, $(\theta, \phi)$,
instead of rapidity space, $(y, \phi)$. This is because experiments usually
operate at the production threshold of the processes of particular interest, so
jets are less boosted than at the LHC.

There are two main algorithms of interest: the Durham Algorithm and the
Generalised \kt\ for \ee.

\subsection{Durham Algorithm}
\label{sec:durham}

For the Durham Algorithm the metric distance between jets $i$ and $j$ is defined as:

$$
d_{ij} = 2 \text{min}(E_i^2, E_j^2) (1 - \cos \theta_{ij})
$$

where $E_{i}$ is the cluster energy and $\theta_{ij}$ is the angular separation between $i$ and $j$.

The reconstruction implementation in \JR\ is then called with

\begin{lstlisting}
    cs = jet_reconstruct(particles; algorithm=Durham)
\end{lstlisting}

No additional parameters are required. In principle the Durham algorithm
proceeds until all clusters are merged to a single jet, but actual analysis will
utilise an exclusive jet cut, for the number of jets of interest.

\subsection{Generalised \kt\ for \ee}
\label{sec:getktee}

For the Generalised \kt\ for \ee\ algorithm, the distance metrics are:

$$
d_{ij} = \text{min}(E_i^{2p}, E_j^{2p}) \frac{1 - \cos \theta_{ij}}{1 - \cos R}
\quad ; \quad
d_{iB} = E_i^{2p}
$$

For a power value $p$ and a radius parameter $R$. This means jets are finalised
when no clusters are found within an angular distance $R$, when $R<\pi$ (this is
very similar to the behaviour of the $pp$ algorithms). In \JR\ we follow the
Fastjet prescription that for $R>\pi$ the denominator is replaced by $3+\cos
R$~\cite{fastjetmanual}.

In the case when $p=1$ and $\pi < R < 3\pi$ the clustering sequence is identical
to the Durham Algorithm, save for a normalisation factor of 2.

\subsection{Implementation Details}
\label{eeimplementation}

For an optimal implementation of the \ee\ algorithms we introduce a new Julia
structure to represent a jet to be reconstructed in $(\theta, \phi)$ space, an \texttt{EEjet}.

This structure mainly differs from the \texttt{PseudoJet} used for $pp$
reconstruction in that the cached values are optimised for the different
reconstruction scheme. Both \texttt{EEjet} and \texttt{PseudoJet} are subtypes
of the abstract type \texttt{FourMomentum}, which allows us to parameterise the
\texttt{ClusterSequence} on the jet type, thus benefiting from a type specific
implementation at runtime, with generic code for the \texttt{ClusterSequence}, viz.

\begin{lstlisting}
    struct ClusterSequence{T <: FourMomentum}
        ...
        jets::Vector{T}
        ...
    end
\end{lstlisting}

As the algorithm is being executed additional compact arrays, that track
parameters used by the calculation (e.g., the nearest neighbour active cluster,
the angle to the nearest neighbour, the fractional momentum in the $x$, $y$ and
$z$ directions, etc.), are maintained for computational efficiency. We use the
Julia package \texttt{StructArrays.jl}~\cite{structarrays}, which allows us to
maintain an ergonomic \emph{array of structures} interface, but underlying this
is a computationally efficient \emph{structure of arrays}, which gives excellent
performance, as shown in Section \ref{sec:eeperf}.

\section{$pp$ and \ee Performance}
\label{sec:performance}

We have benchmarked the performance of \JR\ v0.4.3 against Fastjet
v3.4.3\footnote{The benchmark machine used was an AMD Ryzen 7, 5700G 3.8GHz (8
cores, plus HT), 32GB RAM, running AlmaLinux 9.4. Julia v1.11.1 was used for
\JR\ and Fastjet was compiled with gcc 11.4.1 using \texttt{-O2}. Benchmark runs
are repeated 32 times and are stable to 1\%.}. Pythia was used to generate input
events at different cluster densities. Each sample consisted of 100 events and
where the average density varied from 43 to 632, for $pp$ algorithms and from 43 to
65 for \ee\ algorithms.

\subsection{$pp$ Algorithms}

\begin{figure}[th]
    \begin{center}
        \includegraphics[width=0.8\linewidth]{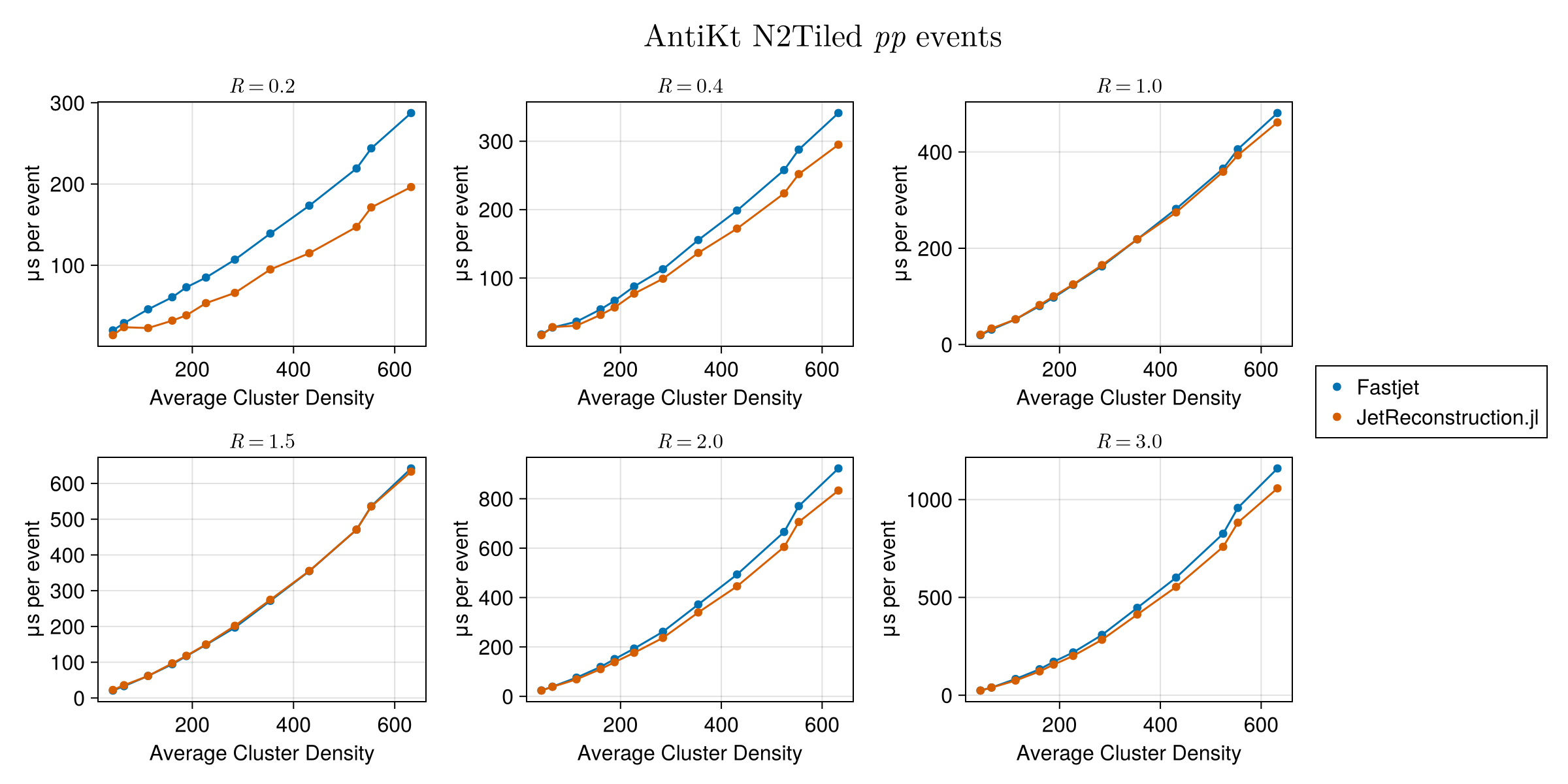}
        \caption{Jet reconstruction time for \akt\ jet reconstruction at different values of $R$, using the \texttt{N2Tiled} strategy. Reconstruction time for Julia and Fastjet are plotted against the average cluster density of different samples.}
        \label{fig:antiktTiledResults}
    \end{center}
\end{figure}

\begin{figure}[th]
    \begin{center}
        \includegraphics[width=0.8\linewidth]{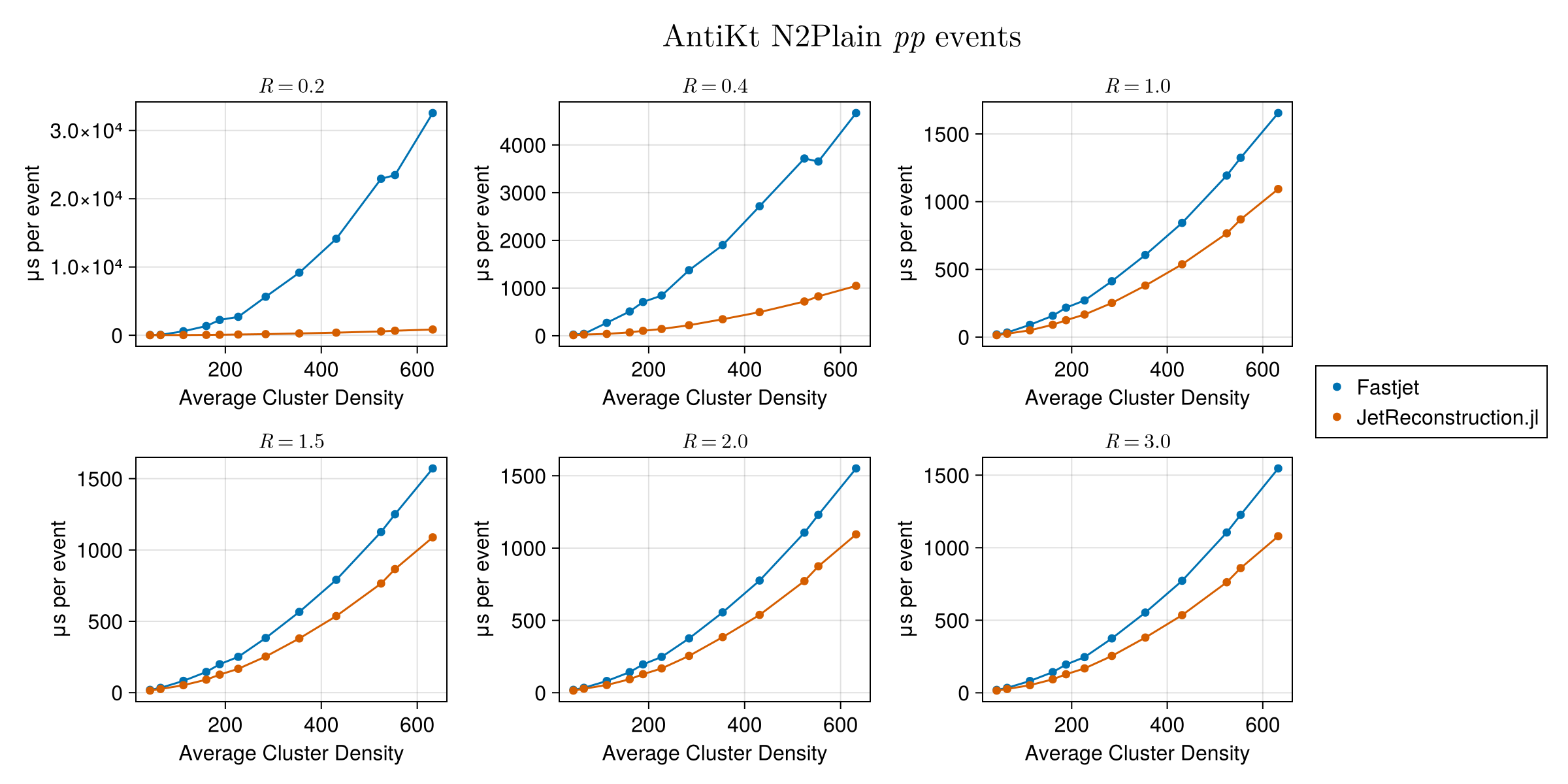}
        \caption{Jet reconstruction time for \akt\ jet reconstruction at different values of $R$, using the \texttt{N2Plain} strategy. Reconstruction time for Julia and Fastjet are plotted against the average cluster density of different samples.}
        \label{fig:antiktPlainResults}
    \end{center}
\end{figure}

The results for \akt\ jet reconstruction are shown in Figures
\ref{fig:antiktTiledResults} and \ref{fig:antiktPlainResults}. The results from
\JR\ are consistently faster than Fastjet. For the more relevant tiled strategy,
which is favoured at $pp$ event densities, Julia is 14-18\% faster, depending on
the particular algorithm chosen, as seen in Table \ref{tab:ppratios}.

\begin{table}[th]
    \begin{center}
        \begin{tabular}{l|ccc}
            \textbf{Algorithm} & \textbf{N2Tiled} & \textbf{N2Plain} & \textbf{N2Plain}, $R>=1.0$ \\
            \hline
            \Akt & 1.14 & 6.46 & 1.49 \\
            Cambridge/Aachen & 1.18 & 6.03 & 1.80 \\
            Inclusive \kt & 1.16 & 6.50 & 1.73 \\
        \end{tabular}
        \caption{Mean ratio of Fastjet reconstruction time to Julia (thus $>1$ indicates Julia is faster) for different $pp$ algorithms, for all values of $R \in (0.2, 0.4, 1, 1.5, 2, 4)$ and all sample cluster densities. As Fastjet's behaviour at $R<1$ for \texttt{N2Plain} is pathological, the mean ratio for $R>=1$ is also given.}
        \label{tab:ppratios}
    \end{center}
\end{table}

\subsection{\ee\ Algorithms}
\label{sec:eeperf}

\begin{figure}[th]
    \begin{subfigure}{0.47\textwidth}
        \begin{center}
            \includegraphics[width=0.8\linewidth]{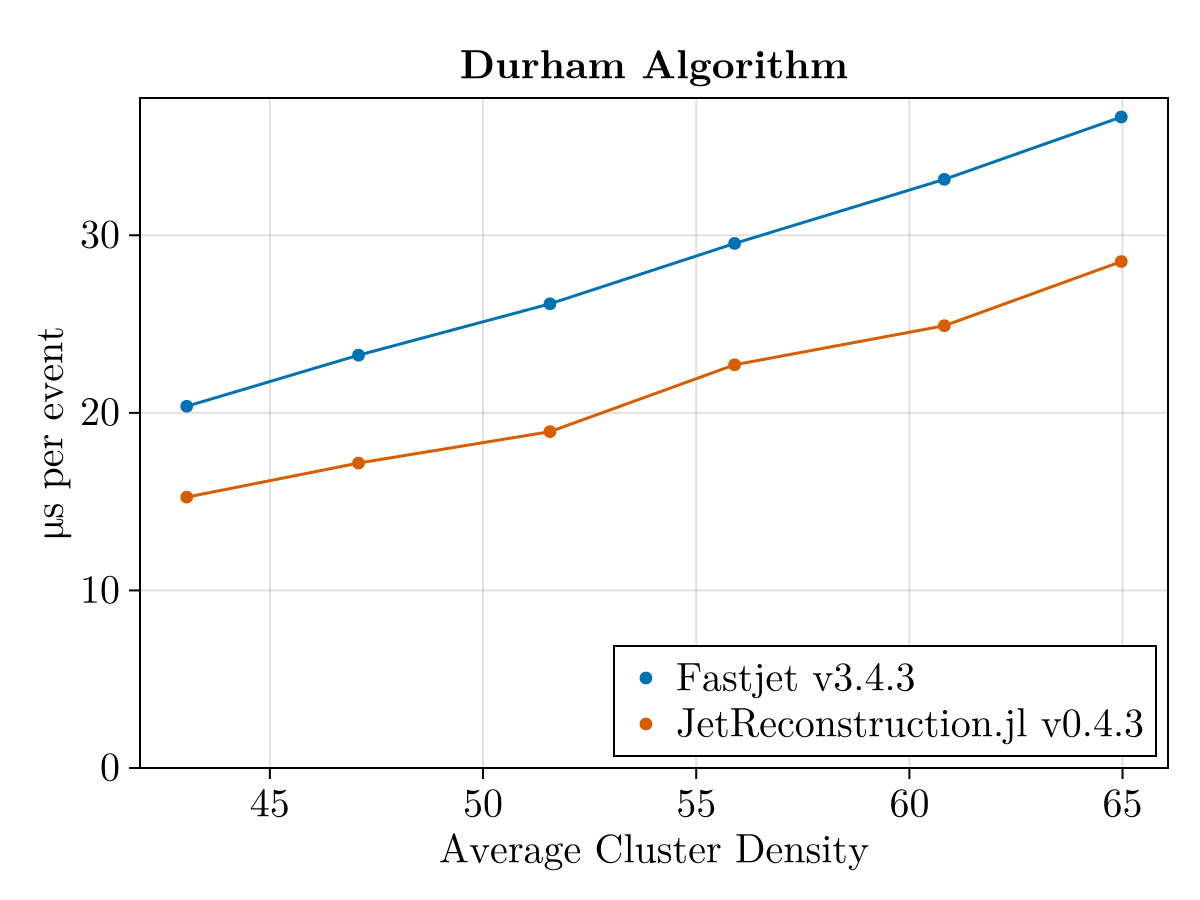}
            \caption{Durham.}
            \label{fig:eeDurham}
        \end{center}
    \end{subfigure}
    \hfill
    \begin{subfigure}{0.47\textwidth}
        \begin{center}
            \includegraphics[width=0.8\linewidth]{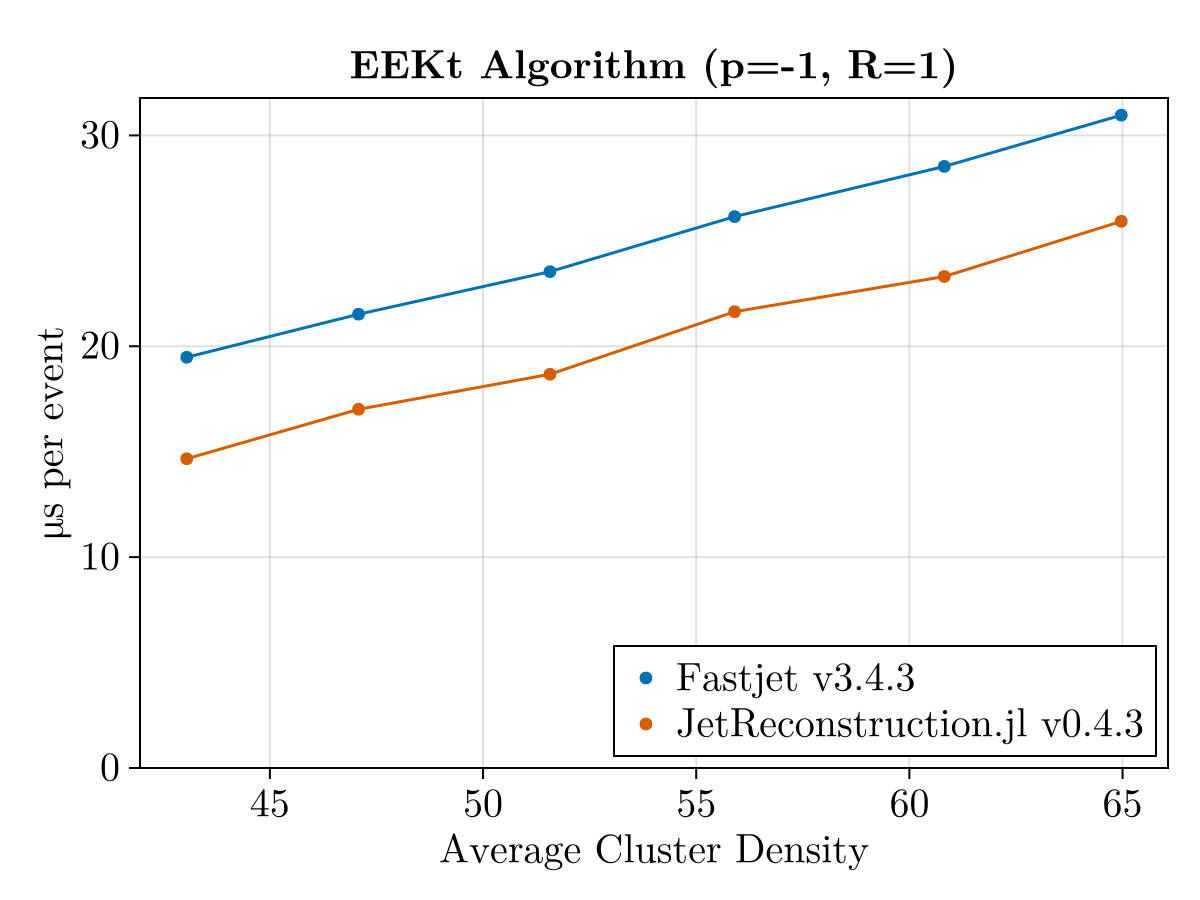}
            \caption{Generalised \kt\ \ee\ $p=-1$, $R=1$.}
            \label{fig:eeKt}
        \end{center}
    \end{subfigure}
    \caption{Jet reconstruction times for \ee\ algorithms. Reconstruction time for Julia and Fastjet are plotted against the average cluster density of different samples.}
    \label{fig:eeperf}
\end{figure}

The performance of the Julia versions of the Durham compared to Fastjet is shown
in Figures \ref{fig:eeDurham} and \ref{fig:eeKt}. The Julia implementation in
\JR\ is consistently faster than Fastjet, by 33\% for the Durham algorithm and,
on average, by 40\% for generalised \kt\ \ee\, as seen in Table \ref{tab:eeratios}.

\begin{table}[th]
    \begin{center}
        \begin{tabular}{l|c}
            \textbf{Algorithm} & \textbf{Runtime Ratio} \\
            \hline
            Durham & 1.33 \\
            Generalised \kt\ \ee & 1.40 \\
        \end{tabular}
        \caption{Mean ratio of Fastjet reconstruction time to Julia (thus $>1$ indicates Julia is faster) for different \ee\ algorithms. The Durham value is averaged over all sample cluster densities. The Generalised \kt\ \ee\ is averaged over all sample cluster densities, $p \in (-1, 0, 1)$ and $R \in (0.2, 0.4, 1, 1.5, 2, 4)$.}
        \label{tab:eeratios}
    \end{center}
\end{table}

\section{Substructure and Taggers}
\label{sec:sstag}

At the LHC jet substructure is a critical component of many physics studies,
probing the internal structure of jets, which can be crucial for distinguishing
between different types of particles and for identifying new physics signals. In
\JR, since v0.4.4, we have implemented several substructure algorithms and taggers that are
commonly used in LHC analyses.

The substructure algorithms \emph{soft
drop}~\cite{Larkoski_2014} and \emph{mass drop}~\cite{Butterworth_2008} are
implemented. To use these taggers, we allow users to define a simple structure
with the relevant tagging parameters, then to call the appropriate tagger
method. The \emph{jet filtering}~\cite{Butterworth_2008} and \emph{jet
trimming}~\cite{Krohn_2010} methods have also been introduced in \JR. These
methods are used to mitigate the effects of pileup and underlying event
contamination in jet reconstruction and are used in a similar way to the
taggers: a simple struct of parameters is defined and the appropriate method
called to obtain a cleaned jet.

The average time to perform the substructure and tagging routines on our
benchmark machine is shown in Table \ref{tab:subtag}. Fastjet is faster
for filtering and trimming, while \JR\ is much faster for the mass drop
algorithm.

\begin{table}[ht]
    \begin{center}
        \begin{tabular}{l|ccc}
            \textbf{Method} & \textbf{\JR} & \textbf{Fastjet} & \textbf{Ratio (Fj/JR)}\\
            \hline
            Jet Filtering & 3.37 & 2.47 & 0.73 \\
            Jet Trimming & 2.80 & 2.26 & 0.81 \\
            Mass Drop & 0.26 & 0.85 & 3.2 \\
            Soft Drop & 2.85 & -- & -- \\
        \end{tabular}
        \caption{Time in $\mu\mathrm{s}$ to perform various substructure and
        tagging operations in \JR\ and in Fastjet (jet filtering is with a
        filter radius of 0.3 and selecting the 3 hardest jets; trimming is uses
        a trim radius and trim fraction of 0.3, with Cambridge/Aachen
        reclustering; mass drop uses $\mu = 0.67$, $y=0.09$; and soft drop
        uses $z_{cut}=0.1$, $\beta=2.0$). The average is taken by performing these
        operations on all inclusive jets above 2GeV in all of our input sample
        files. Note that soft drop is not available in the core Fastjet release.}
        \label{tab:subtag}
    \end{center}
\end{table}

\section{FCCee Jets and EDM4hep}
\label{sec:fccee}

As an example of how to easily support experiment event data models (EDM) in \JR\ we have
introduced support for reading EDM4hep~\cite{Gaede:2022leb} events directly. Like
visualisation, this support takes advantage of the Julia package manager's
concept of \emph{extensions}, which means that the additional code to support
this is not loaded unless the \texttt{EDM4hep.jl} package is present in the
user's environment. Once this is done it is trivial to call the main jet
reconstruction directly using the \texttt{ReconstructedParticles} object from an
EDM4hep ROOT file.

On our benchmark machine, using the Durham algorithm, the jet reconstruction can
reach 24kHz on a single thread.

\section{Conclusions}
\label{sec:conclusions}

\JR\ has been released as a production package in the Julia ecosystem. It is
easy to use and gives better performance than Fastjet in most cases. Feedback
from users has been very positive. We intend to continue to develop that package
in the future, allowing HEP users to benefit from the advantages of Julia in
their analyses that require jet reconstruction.

\sloppy
\raggedright
\bibliography{fast-jet-finding-julia}

\end{document}